\documentclass[review]{elsarticle}

\usepackage{graphicx}% Include figure files
\usepackage{dcolumn}% Align table columns on decimal point
\usepackage{bm}% bold math
\usepackage{epstopdf}
\usepackage{color}
\usepackage{url}
\usepackage{hyperref}
\begin{document}

\begin{frontmatter}

\title{Specific heat ratio effects of compressible Rayleigh-Taylor instability studied by discrete Boltzmann method}

\author[Address1]{Lu Chen}
\author[Address1]{Huilin Lai\corref{mycorrespondingauthor}}
\cortext[mycorrespondingauthor]{Corresponding author}
\ead{hllai@fjnu.edu.cn}
\author[Address2]{Chuandong Lin\corref{mycorrespondingauthor}}
\cortext[mycorrespondingauthor]{Corresponding author}
\ead{linchd3@mail.sysu.edu.cn}
\author[Address1]{Demei Li}

\address[Address1]{College of Mathematics and Statistics, FJKLMAA, Center for Applied Mathematics of Fujian Province (FJNU), Fujian Normal University, Fuzhou 350117, China.}
\address[Address2]{Sino-French Institute of Nuclear Engineering and Technology, Sun Yat-sen University, Zhuhai 519082, China.}

\begin{abstract}
Rayleigh-Taylor (RT) instability widely exists in nature and engineering fields. How to better understand the physical mechanism of RT instability is of great theoretical significance and practical value. At present, abundant results of RT instability have been obtained by traditional macroscopic methods. However, research on the thermodynamic non-equilibrium (TNE) effects in the process of system evolution is relatively scarce. In this paper, the discrete Boltzmann method based on non-equilibrium statistical physics is utilized to study the effects of the specific heat ratio on compressible RT instability. The evolution process of the compressible RT system with different specific heat ratios can be analyzed by the temperature gradient and the proportion of the non-equilibrium region. Firstly, as a result of the competition between the macroscopic magnitude gradient and the non-equilibrium region, the average TNE intensity first increases and then reduces, and it increases with the specific heat ratio decreasing; the specific heat ratio has the same effect on the global strength of the viscous stress tensor. Secondly, the moment when the total temperature gradient in $y$ direction deviates from the fixed value can be regarded as a physical criterion for judging the formation of the vortex structure. Thirdly, under the competition between the temperature gradients and the contact area of the two fluids, the average intensity of the non-equilibrium quantity related to the heat flux shows diversity, and the influence of the specific heat ratio is also quite remarkable.
\end{abstract}

\end{frontmatter}

%\linenumbers

\section{Introduction}

When a heavy fluid is supported or accelerated by a light fluid, the physical phenomenon of small perturbations at the interface growing with time is called Rayleigh-Taylor (RT) instability. Because of the gravitational field, the light fluid rises as a bubble structure and the heavy fluid sinks as a spike structure. When the amplitude is equal to the wavelength, the disturbance begins to become asymmetric, and then grows nonlinearly, and finally enters the turbulent mixing stage\cite{Dimonte2004}. The RT instability is a typical and fundamental interfacial instability that plays an important role in many fields, such as inertial confinement fusion \cite{Ping2019, Lindl2004, Jacquemot2017}, supernova explosions \cite{Ribeyre2004, Fraschetti2010, Hillebrandt2013}, meteorology \cite{Agee1974, Jiang2013}, and geophysics \cite{Houseman1997, Kaus2007, Ghosh2020}. Therefore, the study of RT instability has important theoretical significance and application value, and has aroused considerable interest and been developed rapidly.

The study of RT instability began with the classical linear stability theory proposed by Rayleigh and Taylor, and they discovery that the amplitude of the disturbance in the early stage has exponential growth law\cite{Rayleigh1882, Taylor1950}. In the past few decades, scholars have proposed many numerical methods to study RT instability, including the monte carlo method \cite{Barber2006}, the direct numerical simulation method \cite{Cook2001,Liang2019}, the monotone integrated large eddy simulation \cite{Youngs2009}, the smooth particle hydrodynamics method \cite{Shadloo2013}, the parallel adaptive wavelet collocation method \cite{Reckinger2015}, the phase-field method \cite{Yang2021}, etc. The above methods provide an effective tool for the study of RT instability. By using these methods, authors studied the effects of different physical quantities on RT instability. Wang et al. studied the influence of interface width on the RT instability under weakly nonlinear conditions \cite{Wang2010}. Wei et al. studied the single-mode RT instability problem at low Atwood (At) numbers \cite{Wei2012}. Yang et al. studied the effect of side-wall boundary on RT instability \cite{Yang2021}. Bian et al. systematically analyzed the influence of the disturbed Reynolds number and At number on the late growth of RT instability \cite{Bian2020}.

The above research enriches our in-depth understanding of the RT instability, but at the same time, these conclusions are mainly focus on the macro level, and do not involve concrete thermodynamic non-equilibrium (TNE) information. In order to have a deeper understanding of the TNE evolution law of RT instability systems, we turn to the Boltzmann equation based on non-equilibrium statistical physics, which can not only provide macroscopic fluid dynamics information, but also mesoscopic non-equilibrium information. In the last three decades, the lattice Boltzmann method (LBM) develops rapidly and becomes an effective tool for researching complex dynamics and kinetic behavior, which is self-adaptive in physical description. The discrete Boltzmann method (DBM) is a variant of the standard LBM. Formally, both DBM and LBM use the discrete Boltzmann equation. But in most literature LBM is regarded as a new scheme to solve partial differential equations, while DBM is a coarse-grained physical model based on non-equilibrium statistical physics. It chooses a research perspective and selects a set of kinetic properties to study the system. Thus, it requires that the corresponding kinetic moments describing those properties to keep values in the model simplification \cite{Xu2012,XuAG2018,Xu2021,XuAG2021}. From the perspective of physical modeling, DBM which describes the TNE effect is approximately equivalent to a continuous fluid model and a coarse-grained model. In the region where the Navier-Stokes (NS) model is valid, it is equivalent to the NS equations plus a coarse-grained model of TNE. And in the case of the NS failure, it is equivalent to a modified NS equations plus a coarse-grained model of TNE \cite{Xu2012,Gan2015}. In recent ten years, DBM has been widely used in a variety of complex flow systems \cite{Xu2015,Lai2016,Lin2016,Chen2016,Lin2017,Xu2018,Chen2018,Zhangm2018,GAN2018,LiDM2018,Gan2019,Zhang2019,Ye2020,Chen2020,Lin2020}, and it provides an effective tool to study the non-equilibrium effects in RT instability. Lai et al. studied the influence of compressibility on RT instability, and found that compressibility inhibits the evolution of RT instability in the early stage, but promoted the development in the later stage. At the same time, the TNE also increases with the enhancement of compressibility \cite{Lai2016}. Chen et al. studied the effects of viscosity, heat conduction and Prandtl number on RT instability by using a multiple-relaxation-time DBM \cite{Chen2016}. Lin et al. studied the effect of Reynolds numbers on both global non-equilibrium manifestations and the growth rate of the entropy of mixing though the two-component DBM \cite{Lin2017}. Li et al. used the DBM
to simulate the multi-mode RT instability in a compressible fluid system and explore the evolutionary mechanism \cite{LiDM2018}. Ye et al. used the DBM to explore the influence of Knudsen number on the compressible RT instability, and found that the Knudsen number suppresses the development of RT instability, and also enhances both hydrodynamic non-equilibrium (HNE) and TNE strength \cite{Ye2020}. Chen et al. studied the coupled Rayleigh-Taylor-Kelvin-Helmholtz instability system with the multi-relaxation-time DBM, and studied the complex fluid structure and dynamics process by introducing the morphological boundary length and TNE strength \cite{Chen2020}.

Furthermore, the specific heat ratio is one of the most important parameters to describe the thermodynamic properties of fluids, which is closely related to the compressibility of fluids. It is of great theoretical and application value to study the influence of the specific heat ratio on the evolution of complex fluid system. Bernstein et al. found that the evolution rate of RT instability decreases with the specific heat ratio increasing\cite{Bernstein1982}. Fraley et al. presented a solution for the perturbation growth of density discontinuity caused by shock waves at the material interface, and analyzed the interface asymptotic velocities at different specific heat ratios \cite{Fraley1986}. Livescu et al. found that when the specific heat ratios of the upper and lower fluids are different, the growth rate of RT instability is more sensitive to the change of the specific heat ratio of the lower fluid. However, when the At number is large, the growth rate is less affected by the specific heat ratio of the fluid, and then the equilibrium pressure at the interface becomes the main parameter that affects the compressibility of the system \cite{Livescu2004}. Lafay et al. proved that as the compressibility of the two fluids decreases, the linear growth rate increases, and when the compressibility of two fluids is equal, the linear and nonlinear behaviors are opposite \cite{Lafay2007}. He et al. studied the effect of the interface specific heat ratio on the growth rate of RT instability, and found that when the specific heat ratio of the upper heavy fluid is small, it would hinder the growth of RT instability. And the opposite effect will be achieved when the specific heat ratio of the lower fluid is relatively small \cite{He2008}. Xue et al. found that when the specific heat ratio decreases, the compressibility has a destabilizing effect \cite{Xue2010}. Wang et al. studied the evolution of RT instability related to the low specific heat ratio $(\gamma <5/3)$ in the remnants of the Type Ia supernovae \cite{Wang2011}. Hu et al. studied the impact of the specific heat ratio on the piston effect by simulating eight near-critical fluids \cite{Hu2016}. Zhao et al. studied the bubble growth in compressible RT and RM instabilities, and found that in the RT instability, the adiabatic index and density of the upper fluid increase the amplitude and velocity of the bubble, but reduce the radius of curvature in the early stage, while the influence of the adiabatic index and density of the lower fluid is exactly the opposite\cite{Zhao2020}. These studies have achieved fruitful results and improved our understanding of the physical laws of compressible RT instability.

However, the above research results mainly focus on the description of macroscopic HNE behaviors, and do not contain TNE effect at the intermediate scale. In this paper, the DBM is used to study the effects of the specific heat ratio on RT instability, and both TNE and HNE effects are discussed and analyzed. Furthermore, the non-equilibrium laws of mesoscopic dynamics are presented, which are helpful for us to understand the evolutionary process of multi-scale interface instability. The remaining structure of this paper is as follow: Section \ref{Methodology} introduces the DBM and the details of non-equilibrium quantity extraction. Section \ref{Numerical Simulations} displays the numerical simulation. The effects of the specific heat ratio on RT instability and its physical mechanism are analyzed in detail. Finally, a brief conclusion is given in Section \ref{Conclusions}.

\section{Discrete Boltzmann model}\label{Methodology}

In this work, the discrete Boltzmann equation with the Bhatnagar-Gross-Krook collision term \cite{Zhang2018, Tamura2011, Gan2013} can be described by:
\begin{equation}\label{e5}
\frac{{\partial {f_i}}}{{\partial t}} + {\textbf{v}_i}\cdot\frac{{\partial {f_i}}}{{\partial {\textbf{r}}}} - \frac{{{\textbf{a}}\cdot({\textbf{v}_i} - \textbf{u})}}{{T}}f_i^{eq} =  - \frac{1}{\tau }({f_i} - f_i^{eq}),
\end{equation}
where $f_i$ and $f_i^{eq}$ are the discrete forms of the velocity distribution function and the Maxwellian distribution function, respectively, $t$ is the time, $\textbf{v}_i$ the discrete velocity, $i (=1, 2, \cdots ,N)$ the discrete velocity direction, $\textbf{r}$ the space coordinates, $\textbf{a}$ the external force, $\textbf{u}$ the macroscopic velocity, $T$ the fluid temperature, and $\tau$ the relaxation time.

It can be easily proved that, by means of the Chapman-Enskog multi-scale expansion, the compressible NS equations can be recovered by this model in the hydrodynamic limit:
\begin{eqnarray}\label{e6}
\left\{
\begin{array}{ll}
\frac{\partial \rho}{\partial t} + \nabla \cdot (\rho \textbf{u}) = 0, &  \\[10pt]
\frac{\partial}{\partial t}(\rho \textbf{u}) + \nabla \cdot (\rho \textbf{u}\textbf{u} + p\textbf{I} + \textit{\textbf{P}}) = \rho \textbf{a}, &  \\[10pt]
\frac{\partial}{\partial t} \Big[\rho \Big(e + \frac{|\textbf{u}|^2}{2}\Big)\Big]+ \nabla \cdot \Big[\rho\textbf{u}\Big(e + T + \frac{|\textbf{u}|^2}{2} \Big) + \textit{\textbf{J}} + \textit{\textbf{P}} \cdot \textbf{u} \Big] \\[10pt]
= \rho \textbf{a} \cdot \textbf{u}, &
\end{array}
\right.
\end{eqnarray}
here, $\rho$ denotes the fluid density, $\textit{\textbf{P}}$ and $\textit{\textbf{J}}$ are quantities related to viscous stress and heat flux respectively, and the specific forms are:
\begin{equation}\label{e7}
\textit{\textbf{P}} = -\mu\Big[\nabla \textbf{u} + (\nabla \textbf{u})^T - \frac{2}{D+n}(\nabla \cdot \textbf{u})\textbf{I}\Big],
\end{equation}
\begin{equation}\label{e8}
\textit{\textbf{J}} = -\kappa \nabla T,
\end{equation}
and $e = [(D+n)/2]T$ is the internal energy per unit mass of the system, $\kappa = (D+n+2)\mu/2$ the thermal conductivity coefficient, $\mu = \tau\rho T = \tau p$ the dynamic viscosity coefficient, and $D$ the space dimension, $n$ the number of additional degrees of freedom introduced except translational freedom. In order to obtain a macroscopic result consistent with the compressible NS equations in the continuous limit condition, the local equilibrium distribution function in the discrete Boltzmann equation needs to satisfy the following seven kinetic moment relations:
\begin{eqnarray}\label{e9}
\sum\limits_{i}f_{i}=\sum\limits_{i}f_{i}^{eq}=\rho,
\end{eqnarray}%
\begin{eqnarray}\label{e10}
\sum\limits_{i}f_{i}\textbf{v}_i
=\sum\limits_{i}f_{i}^{eq}\textbf{v}_i=\rho \textbf{u},
\end{eqnarray}%
\begin{eqnarray}\label{e11}
\sum\limits_{i}f_{i}(\textbf{v}_{i} \cdot \textbf{v}_{i} + \eta_{i}^2)
=\sum\limits_{i}f_{i}^{eq}(\textbf{v}_{i} \cdot \textbf{v}_{i} + \eta_{i}^2)
=\rho \big[(D+n)T+\textbf{u} \cdot \textbf{u}\big],\nonumber\\
\end{eqnarray}%
\begin{eqnarray}\label{e12}
\sum\limits_{i}f_{i}^{eq}\textbf{v}_{i} \textbf{v}_{i}
=\rho (T\mathbf{I} +\textbf{u}\textbf{u}),
\end{eqnarray}%
\begin{eqnarray}\label{e13}
\sum\limits_{i}f_{i}^{eq}(\textbf{v}_{i} \cdot \textbf{v}_{i}+\eta_{i}^2)\textbf{v}_{i}
=\rho \textbf{u}\big[(D+n+2)T+\textbf{u} \cdot \textbf{u}\big],
\end{eqnarray}%
\begin{eqnarray} \label{e14}
\sum\limits_{i}f_{i}^{eq}\textbf{v}_{i}\textbf{v}_{i}\textbf{v}_{i}
=\rho\big[T(\textbf{u}_\alpha \textit{\textbf{e}}_\beta \textit{\textbf{e}}_\chi \delta_{\beta\chi}
+\textit{\textbf{e}}_\alpha\textbf{u}_\beta\textit{\textbf{e}}_\chi \delta_{\alpha\chi}
+\textit{\textbf{e}}_\alpha\textit{\textbf{e}}_\beta \delta_{\alpha\beta} \textbf{u}_\chi)\nonumber\\
+\textbf{u}\textbf{u}\textbf{u}\big],\qquad
\end{eqnarray}%
\begin{eqnarray}\label{e15}
\sum\limits_{i}f_{i}^{eq}(\textbf{v}_{i} \cdot \textbf{v}_{i}+\eta_{i}^2)\textbf{v}_{i}\textbf{v}_{i}
=\rho T \big[(D+n+2)T +\textbf{u} \cdot \textbf{u}\big]\textit{\textbf{e}}_\alpha\textit{\textbf{e}}_\beta \delta_{\alpha\beta}  \nonumber\\
+\rho \textbf{u}\textbf{u}\big[(D+n+4)T+\textbf{u}\cdot\textbf{u}\big].\qquad
\end{eqnarray}%
In fact, the above seven kinetic moment relations can be expressed in a matrix form as below:
\begin{equation}\label{e16}
\textbf{C} \cdot \textbf{f}^{eq} = \hat{\textbf{f}}^{eq},
\end{equation}
where
\begin{equation}\label{e17}
\textbf{f}^{eq} = [f_1^{eq}, f_2^{eq}, \cdots , f_N^{eq}]^{\rm{T}},
\end{equation}
\begin{equation}\label{e18}
\hat{\textbf{f}}^{eq} = [\hat{f}_1^{eq}, \hat{f}_2^{eq}, \cdots , \hat{f}_N^{eq}]^{\rm{T}},
\end{equation}
and $\textbf{C}$ is the coefficient matrix concerning the discrete velocity $\textbf{v}_i$ and the free parameter $\eta_i$. To solve $\textbf{f}^{eq}$ requires that the coefficient matrix should have at least $16\times16$ elements, namely, the velocity model should contain at least 16 discrete velocities. At this point, if $\textbf{C}$ is an invertible matrix, Eq. \ref{e16} leads to:
\begin{equation}\label{e19}
\textbf{f}^{eq} = \textbf{C}^{-1} \cdot \hat{\textbf{f}}^{eq}.
\end{equation}

\begin{figure}[tbp]
\center\includegraphics*%
[scale=0.4]{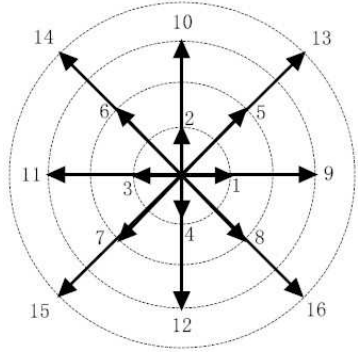}
\caption{Sketch of discrete velocity model.}\label{FIG01}
\end{figure}

In this work, 2-dimensional 16-velocity (D2V16) discrete velocity model is adopted\cite{Lin2016}, as shown in Fig. \ref{FIG01}. In the previous work \cite{Lai2016}, there is only one tunable parameter that controls all values of discrete velocities. In the current work, there are four tunable parameters which indicate magnitudes of four groups of discrete velocities. Compared with the former one, the latter discrete velocity model is more flexible in the selection of discrete velocities, and is beneficial to the robustness and accuracy of the DBM. And the discrete velocity value is obtained by these equations:
\begin{equation}\label{e20}
\textit{\textbf{v}}_i= \left\{
\begin{array}{ll}
v_a\Big[\cos\frac{(i-1)\pi}{2},\sin\frac{(i-1)\pi}{2}\Big], & i=1,\cdots,4, \\[10pt]
v_b\Big[\cos\frac{(2i-1)\pi}{4},\sin\frac{(2i-1)\pi}{4}\Big], & i=5,\cdots,8, \\[10pt]
v_c\Big[\cos\frac{(i-9)\pi}{2},\sin\frac{(i-9)\pi}{2}\Big], & i=9,\cdots,12,\\[10pt]
v_d\Big[\cos\frac{(2i-9)\pi}{4},\sin\frac{(2i-9)\pi}{4}\Big], & i=13,\cdots,16,
\end{array}
\right.
\end{equation}
meanwhile, $\eta_i=\eta_0$ when $i=1,\cdots,4$, and $\eta_i=0$ when $i=5,\cdots,16$.

Both HNE and TNE effects play important roles in non-equilibrium fluid systems. In general, the non-equilibrium information is too complex to be described in detail. Fortunately, the DBM provides an effective tool for studying the non-equilibrium phenomena. Among the seven kinetic moment relations, Eqs. \ref{e9}-\ref{e11} represent the conservation of mass, momentum and energy, respectively, and remain unchanged when the system is moving toward or away from thermodynamic equilibrium. And the TNE information can be extracted from the deviation caused by the $f_{i}$ replacing $f_{i}^{eq}$ in the nonconserved kinetic moments Eqs. \ref{e12}-\ref{e15}.

According to the above ideas, we introduce the kinetic central moments $\textbf{M}_{m,n}^*$ of the velocity distribution function and the equilibrium distribution function \cite{Yan2013},
\begin{eqnarray}\label{e21}
\left\{
\begin{array}{ll}
\textbf{M}_{2}^*(f_i)=\sum\limits_i f_i \textbf{v}_{i}^*\textbf{v}_{i}^*,  \\[10pt]
\textbf{M}_{3}^*(f_i)=\sum\limits_i f_i \textbf{v}_{i}^* \textbf{v}_{i}^*\textbf{v}_{i}^*,\\[10pt]
\textbf{M}_{3,1}^*(f_i)=\sum\limits_i f_i[\textbf{v}_{i}^* \cdot \textbf{v}_{i}^*+\eta_{i}^2]\textbf{v}_{i}^*,\\[10pt]
\textbf{M}_{4,2}^*(f_i)=\sum\limits_i f_i[\textbf{v}_{i}^*\cdot \textbf{v}_{i}^*+\eta_{i}^2]\textbf{v}_{i}^*\textbf{v}_{i}^*,
\end{array}
\right.
\end{eqnarray}
and
\begin{eqnarray}\label{e22}
\left\{
\begin{array}{ll}
\textbf{M}_{2}^*(f_i^{eq})=\sum\limits_i f_i^{eq} \textbf{v}_{i}^*\textbf{v}_{i}^*,   \\[10pt]
\textbf{M}_{3}^*(f_i^{eq})=\sum\limits_i f_i^{eq} \textbf{v}_{i}^* \textbf{v}_{i}^*\textbf{v}_{i}^*,\\[10pt]
\textbf{M}_{3,1}^*(f_i^{eq})=\sum\limits_i f_i^{eq}[\textbf{v}_{i}^* \cdot \textbf{v}_{i}^*+\eta_{i}^2]\textbf{v}_{i}^*,\\[10pt]
\textbf{M}_{4,2}^*(f_i^{eq})=\sum\limits_i f_i^{eq}[\textbf{v}_{i}^*\cdot \textbf{v}_{i}^*+\eta_{i}^2]\textbf{v}_{i}^*\textbf{v}_{i}^*,
\end{array}
\right.
\end{eqnarray}
where $\textbf{v}_{i}^* = \textbf{v}_i - \textbf{u}$, the subscript ``$m, n$'' denotes the reduction of $m$-order tensor to $n$-order tensor. Then, the non-equilibrium quantity can be defined as follows:
\begin{eqnarray}\label{e23}
\boldsymbol{\Delta}_{m,n}^* &=&\textbf{M}_{m,n}^*(f_i)-\textbf{M}_{m,n}^*(f_i^{eq}),
\end{eqnarray}
which $\boldsymbol{\Delta}_{m,n}^*$ describes the information of the thermal fluctuations, including 12 components $\Delta_{2\alpha \beta}^*$, $\Delta_{3\alpha \beta \chi}^*$, $\Delta_{(3,1)\alpha}^*$, $\Delta_{(4,2)\alpha \beta}^*$, $\alpha, \beta, \chi = x$ or $y$. In order to qualitatively analyze the global TNE effect of the system, we define several specific TNE quantities:
\begin{eqnarray}\label{e24}
|\boldsymbol{\Delta}_{2}^*| = |\Delta_{2xx}^*| + |\Delta_{2xy}^*| + |\Delta_{2yy}^*|,
\end{eqnarray}
\begin{eqnarray}\label{e25}
|\boldsymbol{\Delta}_{3,1}^*| = |\Delta_{3,1x}^*| + |\Delta_{3,1y}^*|,
\end{eqnarray}
\begin{eqnarray}\label{e26}
|\boldsymbol{\Delta}_{4,2}^*| = |\Delta_{4,2xx}^*| + |\Delta_{4,2xy}^*| + |\Delta_{4,2yy}^*|,
\end{eqnarray}
\begin{eqnarray}\label{e27}
|\boldsymbol{\Delta}_{3}^*| = |\Delta_{3xxx}^*| + |\Delta_{3xxy}^*| + |\Delta_{3xyy}^*| + |\Delta_{3yyy}^*|.
\end{eqnarray}
And the total TNE quantity can describe the degree of deviation from the equilibrium of the fluid system:
\begin{eqnarray}\label{e28}
|\boldsymbol{\Delta}^*| = |\boldsymbol{\Delta}_{2}^*| + |\boldsymbol{\Delta}_{3,1}^*| + |\boldsymbol{\Delta}_{4,2}^*| + |\boldsymbol{\Delta}_{3}^*|.
\end{eqnarray}

Moreover, the global average TNE strength in the system is defined as:
\begin{eqnarray}\label{e29}
\overline{D} = \frac{1}{L_xL_y}\int_0^{L_x}\int_0^{L_y} |\boldsymbol{\Delta}^*| dxdy,
\end{eqnarray}
the global average viscous stress tensor strength in the system is defined as:
\begin{eqnarray}\label{e30}
\overline{D}_{2} = \frac{1}{L_xL_y}\int_0^{L_x}\int_0^{L_y} |\boldsymbol\Delta_{2}^*| dxdy,
\end{eqnarray}
and the global average heat flux strength in the system is defined as:
\begin{eqnarray}\label{e31}
\overline{D}_{3,1} = \frac{1}{L_xL_y}\int_0^{L_x}\int_0^{L_y} |\boldsymbol\Delta_{3,1}^*| dxdy,
\end{eqnarray}
where $L$ represents the boundary length of the simulation region.

\section{Numerical Simulations}\label{Numerical Simulations}

In this section, numerical simulations are carried out in a two-dimensional rectangular region $\Omega = [-d/2, d/2] \times [-2d, 2d]$. The system with upper and lower layers of fluid is placed in a gravitational field with constant gravitational acceleration, and a cosine disturbance $y_c(x) = y_0\cos(kx)$ is applied at the interface between the two fluids, where $y_0 = 0.05d$, $k = 2\pi/\lambda$ denotes wave number, and $\lambda = d$ the perturbance wavelength. At the initial moment, the system satisfies the static equilibrium condition:
\begin{eqnarray}\label{e32}
\partial_y p_0(y) = -g\rho_0(y),
\end{eqnarray}
where $g$ indicates the magnitude of gravitational acceleration. And the initial conditions of the unstable system are:
\begin{equation}\label{e33}
\left\{
\begin{array}{l}
T_0(y)=T_u,~\rho_0(y)=\frac{p_0}{T_u}\exp{\big[\frac{g}{T_u}\big(2d-y\big)\big]},~y>y_c(x), \\[8pt]
T_0(y)=T_b,~\rho_0(y)=\frac{p_0}{T_b}\exp\big[\frac{g}{T_u}\big(2d-y_c(x)\big)
\\[8pt]
-\frac{g}{T_b}\big(y-y_c(x)\big)\big],~y<y_c(x),
\end{array}
\right.
\end{equation}
where $p_0$ represents the pressure at the top of the system, $T_u$ and $T_p$ express the initial temperatures of the upper and lower fluid, respectively. At the same time, the pressure beside the interface of the two fluids should be consistent, which can be known from the equation of state of ideal gas:
\begin{equation}\label{e34}
\rho_u T_u=\rho_b T_b,
\end{equation}
where $\rho_u$ and $\rho_b$ are the densities of the upper and lower fluids near the interface, respectively.

In the simulation process, a computing grid $N_x \times N_y = 256 \times 1024$ is used. The top and bottom boundaries are dealt with solid wall boundary conditions, and the left and right boundaries are dealt with periodic boundary conditions, the time and space steps are $\Delta t = 2 \times 10^{-5}$, and $\Delta x = \Delta y = 5\times10^{-4}$, respectively. The initial pressure at the top of the system is $p_0 = 1.0$, and the relaxation time $\tau = 4 \times 10^{-5}$. The temperatures at the upper and lower parts are $T_u = 1.0$ and $T_p = 4.0$, respectively. And the other parameters are $(v_a, v_b, v_c, v_d) = (0.9, 0.9, 4.2, 4.2)$, $\eta_0 = 15$, $a_x = 0.0$, $a_y = -g = -1.0$.

Next, we will focus on the influence of the specific heat ratio $\gamma$ on the RT instability. The specific heat ratio is one of the most important parameters to describe the thermodynamic properties of gases. And the expression of $\gamma$ is as follows:
\begin{eqnarray}\label{e35}
\gamma = \frac{n+4}{n+2}.
\end{eqnarray}

\begin{figure}[tbp]
\center\includegraphics*%
[scale=0.17]{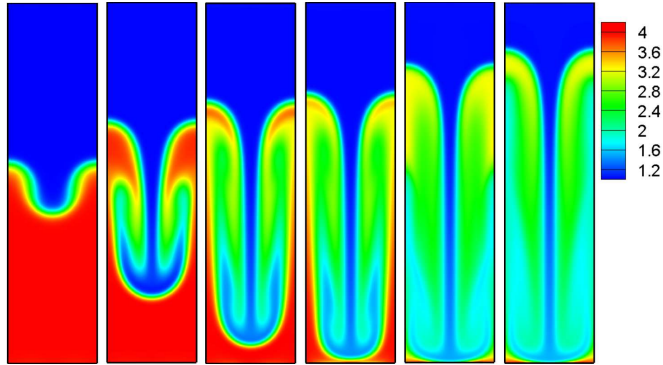}
\caption{Temperature contours at different time instants (from left to right) $t$ = 0.8, 1.6, 2, 2.14, 2.68, and 3, respectively.}\label{FIG02}
\end{figure}

\begin{figure*}
\begin{center}
\includegraphics[bbllx=200pt,bblly=0pt,bburx=592pt,bbury=477pt,width=0.4\textwidth]{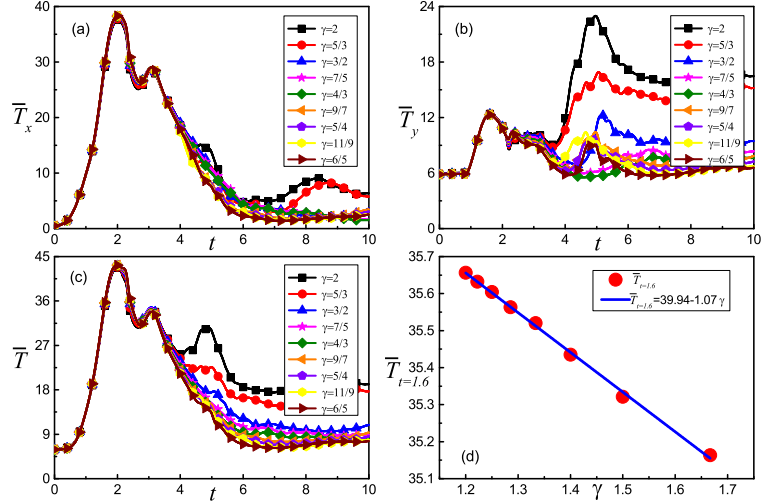}
\end{center}
\caption{Evolution of global average temperature gradients with various specific heat ratios: (a) the global average temperature gradient in the $x$ direction, (b) the global average temperature gradient in the $y$ direction, (c) the global average temperature gradient, (d) the relationship between the gradient value at $t=1.6$ and the specific heat ratio.}\label{FIG03}
\end{figure*}

In order to study the effect of $\gamma$, we adjust the number of extra degrees of freedom $n$ to change $\gamma$. Here, we choose $n=0,1,2,\cdots,8$, correspondingly $\gamma =2$, 5/3, 3/2, 7/5, 4/3, 9/7, 5/4, 11/9, 6/5. In order to observe the changes of interface length and width more clearly, the spatial distribution diagram of temperature is presented with $\gamma =7/5$, as shown in Fig. \ref{FIG02}. Firstly, the global average temperature gradients in fluid system are discussed. The formula for calculating the global average temperature gradients are shown below:
\begin{equation}\label{e33}
\overline{T_x}=\frac{1}{L_x L_y} \int_0^{L_x}\int_0^{L_y}|\nabla_x T|dxdy,
\end{equation}
\begin{equation}\label{e33}
\overline{T_y}=\frac{1}{L_x L_y}\int_0^{L_x}\int_0^{L_y} |\nabla_y T|dxdy,
\end{equation}
\begin{equation}\label{e33}
\overline{T}=\frac{1}{L_x L_y}\int_0^{L_x}\int_0^{L_y} |\nabla T|dxdy.
\end{equation}

Figures \ref{FIG03} (a)-(b) delineate the evolution of the global average temperature gradients in $x$ and $y$ directions, respectively, and Fig. \ref{FIG03} (c) illustrates the global average temperature gradient of the fluid system over time. It can be seen from Fig. \ref{FIG03} (a) that the global average temperature gradients in $x$ direction corresponding to different $\gamma$ values have roughly the same changing trend. Take the case with $\gamma = 7/5$, $\overline{T_x}$ gradually rises and presents an exponential growth state, and at $t = 2.06$, the $\overline{T_x}$ no longer increases and then shows a downward trend. At $t = 2.68$, $\overline{T_x}$ shows a short-term upward trend, and this state continues until $t = 3$, after that $\overline{T_x}$ starts decreasing all the way down.

The physical reason for this phenomenon is that, at the initial stage, the distance between the bubble and the spike increases exponentially, and the change of the temperature in the $x$ direction is concentrated in the middle range between the spike and the bubble, so the temperature gradient shows an exponential growth trend in the early stage, at this time, a large number of high-temperature and low-density fluids at the bottom have risen to low-temperature and high-density flow fields, which reduces the temperature at the bottom and increases the temperature at the top. In the $x$ direction, the temperature at the bottom generally presents a state of high temperature fluid on both sides and low temperature fluid in the middle. When the spike is about to touch the bottom, the mixing of two fluids further increases and even tends to be saturated, this makes the temperature in the system tend to be consistent, leading to the decrease in the global average temperature gradient. At the same time, with the increase of $\gamma$, the values of $\overline{T_x}$ are roughly the same, but there is a big difference near the peak value. Due to the disordered development of the system, the physical mechanism becomes more complex in the latter stage.

However, the change of the global average temperature gradient in the $y$ direction is completely different from the one in the $x$ direction. As can be seen from Fig. \ref{FIG03} (b), $\overline{T_y}$ keeps constant from $t = 0$ to $t = 0.85$, and increases from $t = 0.85$ to $t = 1.6$. After $t = 1.6$, $\overline{T_y}$ begins to show a downward trend until the spike arrive at the bottom. Then, the RT system enters into the chaotic stage, and the evolution of $\overline{T_y}$ becomes more complex. Throughout the evolution, an interesting phenomenon can be observed: $\overline{T_y}$ always keeps constant during the initial period. This is because, in the early stage, the change of the temperature is concentrated in the front of the spike and bubble, and the change is uniform, while in other regions, the change of the temperature are negligible, which makes $\overline{T_y}$ maintain constant in the early stage. As the system evolves, some small vortex structures appear due to the shear stress in the fluid system, and the temperature no longer varies monotonously in the vertical direction, which leads to the gradual increase of $\overline{T_y}$. Therefore, the moment when $\overline{T_y}$ is out of the fixed value can be taken as the criterion to judge the formation of the vortex structure. Fig. \ref{FIG03} (c) illustrates the change curve of the temperature gradient of the fluid system. It can be observed that the maximum of $\overline{T_x}$ is twice that of $\overline{T_y}$. The integral absolute value of temperature gradient in the $x$ direction ($y$ direction) is associated with the changing trend of temperature gradient along the $x$ direction ($y$ direction) and the vertical (horizontal) length of the computational region. On the one hand, within the mixing area between the hot and cold media, the value of temperature gradient first decreases then increase in the $x$ direction, and the value of temperature gradient reduces monotonously in the $y$ direction. On the other hand, the vertical length $L_y$ is greater than the horizontal length $L_x$. In other words, the area where the temperature gradient in the $x$ direction ($y$ direction) deviates from zero is larger (smaller). Consequently, the maximum of temperature gradient in the $x$ direction is larger than that in the $y$ direction. Moreover, with the increase of $\gamma$, the global average temperature gradient shows an decreasing trend. This is because for a larger specific ratio, the morphological structure of the fluid becomes more complex, and the temperature field has larger spatial changes. Furthermore, it can be found that the value of $\overline{T}$ at $t = 1.6$ and $\gamma$ have the following functional relationship: $\overline{T}_{t=1.6} = 39.94-1.07\gamma$.

\begin{figure}[tbp]
\center\includegraphics*%
[scale=0.18]{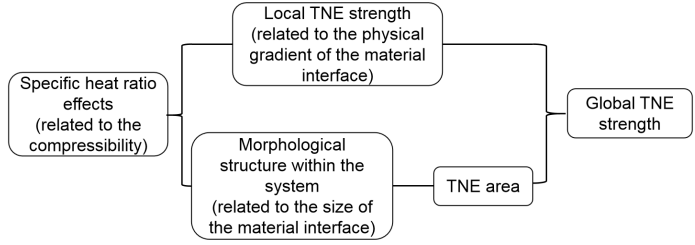}
\caption{Schematic diagram of the ways that the specific heat ratio effects affect the global TNE intensity.}\label{FIG04}
\end{figure}

\begin{figure}[tbp]
\center\includegraphics*%
[scale=0.28]{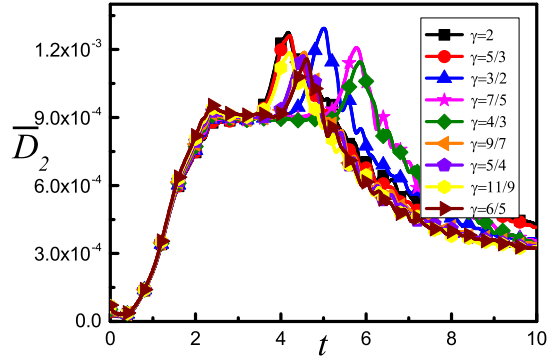}
\caption{Evolution of the global average strength $\overline{D}_2$.}\label{FIG05}
\end{figure}

Next the TNE behaviors of the RT instability are investigated, and then the change of different TNE quantities are analyzed. The ways the specific heat ratio effects affect the global TNE intensity are shown in Fig. \ref{FIG04}. Figure \ref{FIG05} shows the evolution of the global average intensity $\overline{D}_2$ of the non-equilibrium quantity $\Delta_{2\alpha \beta}^*$ over time. It can be found that, $\overline{D}_2$ increases first and then decreases. Furthermore, $\overline{D}_2$ increases with the decrease of $\gamma$. To explain the trend of $\overline{D}_2$ in detail, we take $\gamma =7/5$ as an example, it can be found that $\overline{D}_2$ increases rapidly at the early period, and with the spike reaches bottom at $t = 2.14$, $\overline{D}_2$ first shows a down trend and then gradually increases, and at $t = 5.8$, it reaches the peak, after that $\overline{D}_2$ starts to oscillate less. Next, the effect of $\gamma$ on $\overline{D}_2$ is analyzed. When $t < 2.14$, there is no significant difference in fluid structure between different examples. At this point, with the increase of $\gamma$, the compressibility decreases and the TNE strength decreases, which is consistent with our previous work. At $t > 2.14$, the morphological structure of the system generates great difference, and the complexity of the morphological structure makes the evolution of $\overline{D}_2$ have no obvious rule.

\begin{figure*}[tbp]
\begin{center}
\includegraphics[bbllx=200pt,bblly=0pt,bburx=592pt,bbury=477pt,width=0.4\textwidth]{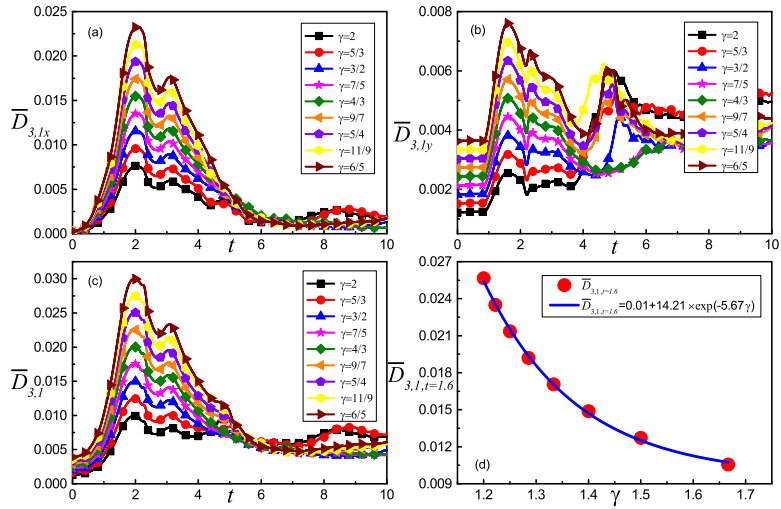}
\end{center}
\caption{TNE quantities $\overline{D}_{3,1x}$, $\overline{D}_{3,1y}$ and $\overline{D}_{3,1}(=|\overline{D}_{3,1x}|+|\overline{D}_{3,1y}|)$ versus time under different specific heat ratios. (d) the fitting curve of the value of $\overline{D}_{3,1}$ at $t=1.6$ and $\gamma$ as: $\overline{D}_{3,1,t=1.6} = 0.01+14.21
\times\exp(-5.67\gamma)$.}\label{FIG06}
\end{figure*}
Figure \ref{FIG06} shows the evolution of the average intensity and global average intensity of each component of $\Delta_{3,1\alpha}^*$ with time. It can be found from the three subgraphs that $\overline{D}_{3,1x}$, $\overline{D}_{3,1y}$ and $\overline{D}_{3,1}$ all decrease with the increase of $\gamma$. To explain this physical phenomena, we resort to Eq. \ref{e8}. It can be seen that both $\kappa$ and $\nabla T$ vary with $\gamma$ in Eq. \ref{e8}, however, as shown in Fig. \ref{FIG03}, the temperature gradients in different cases are similar, so the change of $\textit{\textbf{J}}$ is mainly controlled by $\kappa$. Therefore, the absolute value of $\textit{\textbf{J}}$ decrease with the increase of $\gamma$, which explains the relationship between each non-equilibrium quantity and $\gamma$. In addition, Fig. \ref{FIG06} (a) shows that when $\gamma =7/5$, $\overline{D}_{3,1x}$ rapidly increases to the maximum at the beginning, and then declines. There are two main physical mechanisms in the whole evolution process: one is the growth rate of the mixing area between hot and cold fluids, and the other is the heat conduction caused by the change of temperature gradients. The increase of the area of contact between two fluids causes the increase of the area of heat exchange. And the increase of the temperature gradients also lead to the increase of the heat exchange. In the early stage, the area of contact between two fluids is increasing and the total temperature gradient is also increasing, so the heat exchange is also increasing. After the spike touches the bottom, $\overline{T_x}$ decreases rapidly. Although the contact area is still increasing, the growth rate becomes slower. At this moment, the change of temperature gradient is dominant. For $t = 2.68-3$, the temperature gradient rises again, so $\overline{D}_{3,1x}$ increases. Later, the contact area between the two fluids does not increase with the development of evolution, and the temperature gradient always plays a decisive role, so $\overline{D}_{3,1x}$ will gradually decrease and remain near a fixed value.

\begin{figure}[tbp]
\center\includegraphics*%
[scale=0.2]{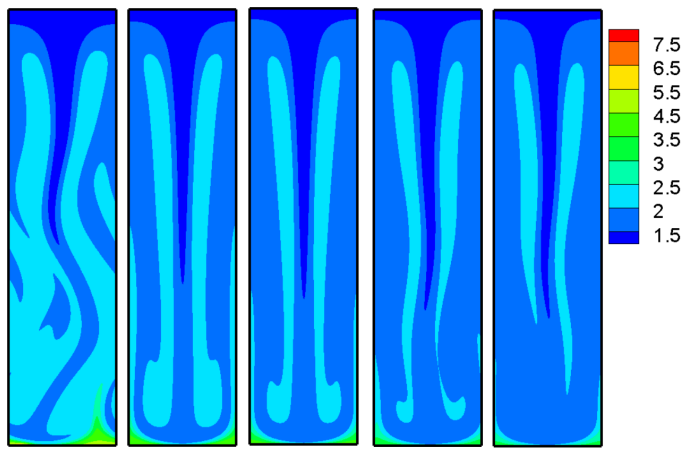}
\caption{Temperature contours corresponding to $\gamma$= 2, 3/2, 4/3, 5/4, 6/5 from left to right with the same time $t = 4$. The color from blue to red indicates the increase of values.}\label{FIG07}
\end{figure}

From Fig. \ref{FIG06} (b), it can be observed that $\overline{D}_{3,1y}$ maintains constant from $t =0$ to $t =0.85$. This is because the total temperature gradient in $y$ direction keeps constant, while the contact area grows slowly. Therefore, the temperature gradient plays a greater role at this period. When $t = 0.85-1.6$,  the contact area of the two fluids increase due to the appearance of the vortex structure, and the total temperature gradient is also rising. As a result, the combined action of the temperature gradient and the increase of the area of contact leads to the increase of $\overline{D}_{3,1y}$. And when $t > 1.6$, $\overline{D}_{3,1y}$ shows a downward trend. With the spike arrives at the bottom, the RT system enters into the chaotic stage, and the evolution of $\overline{D}_{3,1y}$ becomes more complex. Interestingly, as $\gamma$ decreases, $\overline{T_y}$ does not change much, just only a slight difference, but the interaction between the contact area and the temperature gradient has a great influence on the numerical change of $\overline{D}_{3,1y}$. Furthermore, in the later stage, the evolution of $\overline{T_y}$ and $\overline{D}_{3,1y}$ under different $\gamma$ has no obvious rule. In order to explain the inherent physical mechanism, the temperature contour map at $t = 4$ is given by Fig. \ref{FIG07}. It can be observed that there is a great difference in the temperature contour under different $\gamma$ conditions. Therefore, it can be known that the difference in the morphology and structure of the flow field is the main reason for the occurrence of this phenomenon.

The global average TNE intensity is closely related to the macroscopic physical gradients (including temperature, density and velocity gradient) and the area of the non-equilibrium region. The numerical study shows that in the early stage of RT instability, the local physical gradients decrease, such as the local temperature gradient, so the local TNE intensity decreases. However, with the development of RT instability, the increasing area of non-equilibrium region makes the global TNE strength increasingly stronger. That is, the effect of the macroscopic physical gradient and the effect of the non-equilibrium area are competing with each other. In order to facilitate the analysis of the evolution law of global non-equilibrium, the proportion of the non-equilibrium region $S_r$ (i.e., the ratio of non-equilibrium area to total area) is defined. Here, we use the following method to calculate it. Namely, to count the number of points which non-equilibrium intensity is greater than the given threshold and then to compare them with the total number of points to get $S_r$.

\begin{figure*}[tbp]
\begin{center}
\includegraphics[bbllx=200pt,bblly=0pt,bburx=592pt,bbury=477pt,width=0.5\textwidth]{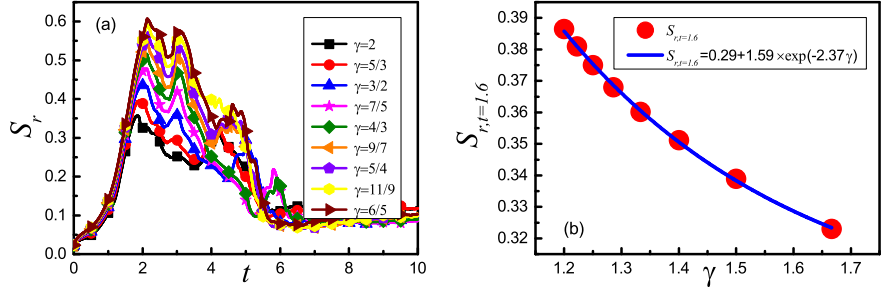}
\end{center}
\caption{(a) evolution of the proportion of the non-equilibrium region $S_r$, (b) relationship between the value of the proportion of the non-equilibrium region $S_r$ at $t=1.6$ and the specific heat ratio $\gamma$.}\label{FIG08}
\end{figure*}

Figure \ref{FIG08} shows the evolution of $S_r$, from which it can be seen that the proportion of the non-equilibrium region increases at first and then decreases. At the beginning, both the non-equilibrium intensity at each point and the non-equilibrium area increase, so that the $S_r$ has an exponential growth trend in the previous period. As time evolves, the fluid system tends to be more balanced, the non-equilibrium strength decreases, and the area of the non-equilibrium region tends to be saturated, which makes $S_r$ gradually decrease. It is further found that there is a relationship between the non-equilibrium area and the specific heat ratio as follows: $S_{r,t=1.6}=0.29+1.59\times\exp(-2.37\gamma)$.

\begin{figure}[tbp]
\center\includegraphics*%
[scale=0.17]{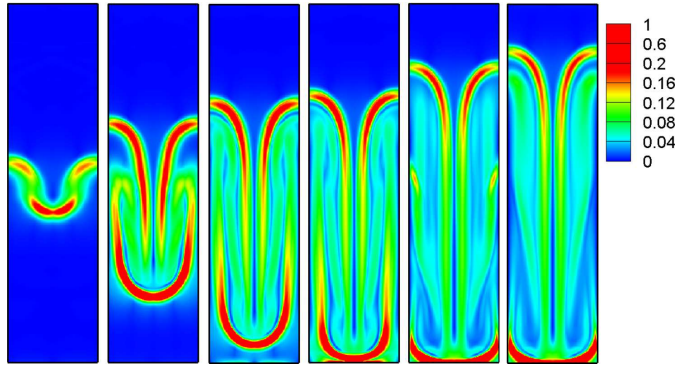}
\caption{Spatial profiles of the global TNE quantity with $\gamma=7/5$ at different time instants (from left to right) $t$ = 0.8, 1.6, 2, 2.14, 2.68, and 3, respectively. The color from blue to red indicates the increase of values. The values of the global TNE quantity is high at the interface, while it is almost zero away from the interface.}\label{FIG09}
\end{figure}

\begin{figure*}[tbp]
\begin{center}
\includegraphics[bbllx=200pt,bblly=0pt,bburx=592pt,bbury=477pt,width=0.5\textwidth]{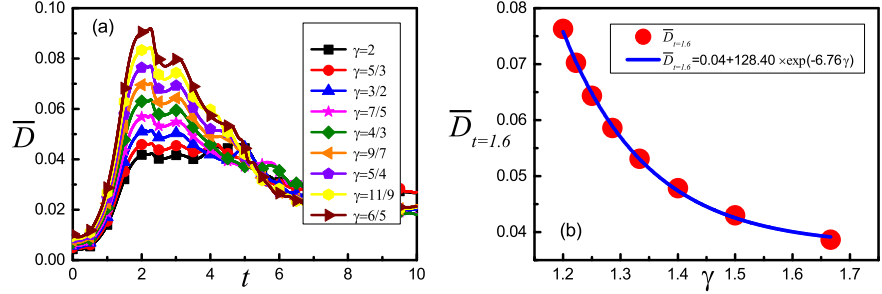}
\end{center}
\caption{(a) the global average TNE intensity $\overline{D}$ evolves with time under different specific heat ratios, (b) the fitting curve of the global average TNE intensity $\overline{D}$ at $t=1.6$ and $\gamma$ as: $\overline{D} = 0.04+128.40
\times\exp(-6.76\gamma)$.}\label{FIG10}
\end{figure*}

Furthermore, the global TNE quantity are shown in Fig. \ref{FIG09}, which can help us to get the spatial distributions of the non-equilibrium region. It can be seen that the TNE intensity which near the front end of the bubble and spike is the highest, this is because the physical gradients near the interface is sharpest. With the evolution of RT instability, the mixing layer gradually elongates, and many small structures appear in the system, leading to the increasing area of the non-equilibrium region. Finally, we discuss the evolution law of the global average TNE intensity $\overline{D}$. Figure \ref{FIG10} shows the evolution of $\overline{D}$ over time. The global average TNE intensity is numerically determined by $\overline{D}_2$, $\overline{D}_{3,1}$, $\overline{D}_3$ and $\overline{D}_{4,2}$, and its intrinsic evolution rules are determined by the physical gradients and the non-equilibrium area of the system. The numerical results show that $\overline{D}$ increases rapidly in the beginning. During this process, the proportion of the non-equilibrium region plays a leading role. In addition, the non-equilibrium area increases with the disturbance amplitude increasing, which leads to the global TNE intensity shows an exponential growth trend. At the time $t = 2.14$, $\overline{D}$ stops growing and starts to decline slowly. This is because the non-equilibrium area shows a downward trend at this time, thus the global non-equilibrium intensity decreases. Meanwhile, it can be seen from Fig. \ref{FIG10} (b) that there is an exponential relationship between the global average non-equilibrium intensity $\overline{D}$ at $t = 1.6$ and the specific heat ratio $\gamma$ as follows: $\overline{D}_{t=1.6} = 0.04+128.40\times\exp(-6.76\gamma)$. Although this work focus on two-dimensional case, the model with flexible specific heat ratio can be applied to pick up information from three-dimensional physical fields. And the non-equilibrium information of three-dimensional RT instability in compressible flows deviating far away from thermodynamic equilibrium deserve further study. But on the other hand, the two-dimensional model does not take into account the geometric effects of the three-dimensional system.

\section{Conclusions}\label{Conclusions}

The DBM has been applied to study the effects of the specific heat ratio on the compressible RT instability. Firstly, two kinds of non-equilibrium quantities related to the viscous stress and heat flux are investigated. It is found that with the increase of the specific heat ratio, the global average strength of the non-equilibrium quantity $\Delta_{2\alpha \beta}^*$ related to the viscous stress decreases, and the overall trend is first increasing and then decreasing. Secondly, with the increase of the specific heat ratio, the total average intensity of the non-equilibrium quantity $\Delta_{3,1\alpha}^*$ related to the heat flux will decrease, and the overall performance is first increases and then decreases. There are two main factors contributing to this phenomenon, i.e., (i) the increasing temperature gradient increases the non-equilibrium strength, (ii) the increase of the area of contact between the two fluids promotes the heat exchange, thus enhancing the non-equilibrium strength. These two factors compete with each other and jointly determine the overall average intensity. Furthermore, the global average thermodynamic non-equilibrium strength $\overline{D}$ is discussed. The numerical simulation results reveal that $\overline{D}$ also increases first and then decreases, and $\overline{D}$ also increases with the decrease of the specific heat ratio. This is because the increasing non-equilibrium area enhances the global non-equilibrium effects, and the decreasing physical gradients weakens the local TNEs. The combined action of the two factors determine the global non-equilibrium strength of the system. These results of the macro and mesoscopic non-equilibrium information contribute to a deeper understanding of the physical mechanisms of the compressible RT instability.

\section*{Acknowledgments}
This work was supported by National Natural Science Foundation of China (under Grant Nos. 51806116, 11875001), Natural Science Foundation of Fujian Provinces (under Grant Nos. 2018J01654).

\appendix

\section{}\label{APPENDIXA}

\begin{figure}[tbp]
\center\includegraphics*%
[scale=0.5]{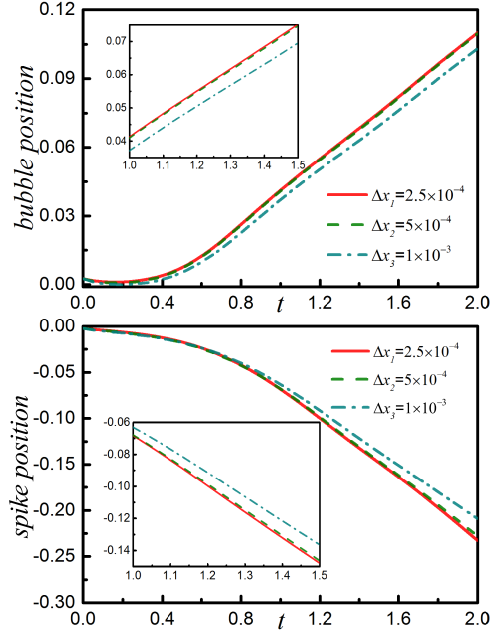}
\caption{The position of the spike and bubble evolves over time with various spatial steps.}\label{FIG11}
\end{figure}

As numerical accuracy should be under consideration, we carry out a grid-independent test of the RT instability with various spatial steps. The RT instability with $n=0$ is simulated in the region of $L_x \times L_y= 0.128\times0.512$. In Fig. \ref{FIG11} the dash-dotted, dashed, and solid lines represent numerical results under spatial steps $\Delta x = \Delta y = 1\times10^{-3}, 5\times10^{-4}$ and $2.5\times10^{-4}$, respectively. And the corresponding meshes are $N_x \times Ny = 128\times512, 256\times1024$ and $512\times2048$, respectively. It can be found that the simulation results are converging with decreasing space steps, and the results with space steps $5\times10^{-4}$ and $2.5\times10^{-4}$ are quite close to each other. Therefore, the space steps $5\times10^{-4}$ and mesh $256\times1024$ are selected by considering the computational cost and accuracy.

\section*{References}

\bibliography{References}

\end{document}